\documentclass{svjour3}

\usepackage{multicol}
\usepackage{color}
\usepackage{xspace}

\usepackage[utf8]{inputenc} 
\usepackage{subfigure}
\usepackage{graphicx,url}
\usepackage{booktabs}
\usepackage{epstopdf}
\usepackage{epsfig}

\begin{document}

\title{Computational Social Scientist Beware: Simpson's Paradox in Behavioral Data}

\author{Kristina Lerman}

\institute{K. Lerman \at
              USC Information Sciences Institute, \\
              Marina del Rey, CA 90292\\
              Tel.: +1-310-913-0091\\
              \email{lerman@isi.edu}
              }

\date{Received: date / Accepted: date}
\maketitle

\begin{abstract} 


Observational data about human behavior is often heterogeneous, i.e., generated by subgroups within the population under study that vary in size and behavior. Heterogeneity predisposes analysis to Simpson's paradox, whereby the trends observed in data that has been aggregated over the entire population may be substantially different from those of the underlying subgroups. I illustrate Simpson's paradox with several examples coming from studies of online behavior and show that aggregate response leads to wrong conclusions about the underlying individual behavior. I then present a simple method to test whether Simpson's paradox is affecting results of analysis.
The presence of Simpson's paradox in social data suggests that important behavioral differences exist within the population, and failure to take these differences into account can distort the studies' findings.

\end{abstract}

\section{Introduction}


The landscape of  social science  changed dramatically in the 21st century, when large volumes of social and behavioral data created the field of computational social science~\cite{Lazer09}. 
While the bulk  of the data is now digital traces of online behaviors, the accelerating instrumentation of our physical spaces 
is opening offline behaviors to analysis.
The new data has vastly expanded the opportunities for discovery in the social sciences~\cite{mcfarland2016sociology}.
Algorithms have mined behavioral data to validate theories of individual decision-making and social interaction~\cite{kleinberg2017human,bond201261} and produce new insights into first principles of human behavior. These insights help to better explain and predict human behavior, and eventually even help policy makers devise more effective interventions to improve wellbeing by steering behaviors towards desirable outcomes, for example, by fostering behaviors that promote healthy habits, reduce substance abuse and social isolation, improve learning, etc.

Computational social scientists, however, are facing challenges, some of which were rarely encountered by past researchers. 
Although behavioral data is usually massive, it is also often extremely sparse and noisy at the individual level. To uncover hidden patterns, 
scientists might choose to aggregate data over the entire population. For example, diurnal cycles of mood (\emph{cf} Fig.~1 in \cite{golder2011diurnal}) and online activity (\emph{cf} Fig.~2 in \cite{Hogg12epj}) only become apparent once the activity of tens of thousands or even millions of people is aggregated.
In the past, when behavioral data came from populations that were carefully crafted to address specific research questions~\cite{mcfarland2016sociology}, aggregation helped improve signal-to-noise ratio and uncover weak effects.
Today, however, the same strategy can lead researchers to wrong conclusions.
The reason for this is that current behavioral data is highly heterogeneous: it is collected from subgroups that vary widely in size and behavior.
Heterogeneity is evident in practically all social data sets and can be easily recognized by its hallmark, the long-tailed distribution. The prevalence of some trait in these systems, whether the number of followers in an online social network, or the number of words used in an email, can vary by many orders of magnitude, making it difficult to compare users with small values of the trait to those with large values.
As shown in this paper, heterogeneity can dramatically distort conclusions of analysis.

Simpson's paradox~\cite{Blyth1972,Norton2015simpson} is one important phenomenon confounding analysis of heterogeneous social data.
According to the paradox, an association observed in data that has been aggregated over an entire population may be quite different from, and even opposite to, associations found in the underlying subgroups.
A notorious example of Simpson's paradox comes from the gender bias lawsuit against UC Berkeley~\cite{Bickel1975}. Analysis of graduate school admissions data seemingly revealed a statistically significant bias against women: a smaller fraction of female applicants were admitted for graduate studies. However, when admissions data was disaggregated by department, women had parity and even a slight advantage over men in some departments. The paradox arose because departments preferred by female applicants have lower admissions rates for both genders.


Simpson's paradox also affects analysis of trends. When measuring how an outcome variable changes as a function of some independent variable, the characteristics of the population over which the trend is measured may change with the independent variable. As a result, the data may appear to exhibit a trend, which disappears or reverses when the data is disaggregated by subgroups~\cite{Alipourfard2018}.
Vaupel and Yashin~\cite{Vaupel85heterogeneity} give several illustrations of 
this effect. For example, a study of recidivism among convicts released from prison showed that the rate at which they return to prison declines over time. From this, policy makers concluded that age has a pacifying effect, with older convicts less likely to commit crimes. In reality, this is not the case. Instead, the population of ex-convicts is composed of two subgroups with nearly constant, but very different recidivism rates. The first subgroup---the ``reformed''---will never commit a crime once released from prison. The other subgroup---the ``incorrigibles''---are highly likely commit a crime. Over time, as ``incorrigibles'' commit offenses and return to prison, there are fewer of them left in the population. Survivor bias changes the composition of the population, creating an illusion of an overall decline in recidivism. As Vaupel and Yashin warn, ``unsuspecting researchers who are not wary of heterogeneity's ruses may fallaciously assume that observed patterns for the population as a whole also hold on the sub-population or individual level.''


To highlight the perils of ignoring Simpson's paradox, I describe several studies of online behavior in which the trends discovered in aggregate data lead to wrong conclusions about behavior. For decision makers and platform designers seeking to use research findings to inform policy, incorrect interpretation can lead to counterproductive choices where a policy thought to enhance some behavior instead suppresses it, or vice-versa. To identify such cases, I present a simple method researchers can use to test for the presence of the paradox in their data. When paradox is confirmed, analysis should be performed on the stratified data that has been disaggregated by subgroups~\cite{Alipourfard2018,Norton2015simpson}.
Testing and controlling for Simpson's paradox should be part of every computational social scientist's toolbox.

\section{Examples of Simpson's Paradox}
Multiple examples of Simpson's paradox have been identified in empirical studies of online behavior. For example, a study of Reddit~\cite{Barbosa2016} found that average comment length decreased over time. However, when data was disaggregated by cohorts based on the year the user joined Reddit, comment length within each cohort increases.
Additional examples of Simpson's paradox are described below.

\paragraph{Exposure Response in Social Media.}

\begin{figure*}[t]\centering
\begin{tabular}{cc}
\includegraphics[height=1.5in]{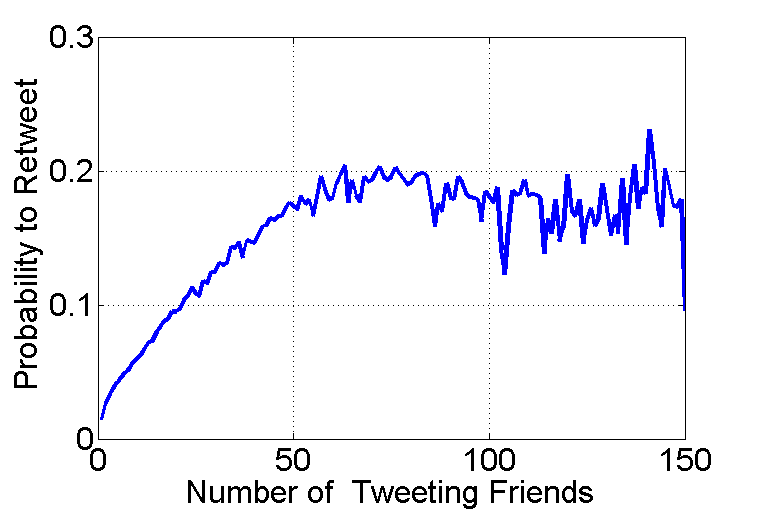}
\includegraphics[height=1.5in]{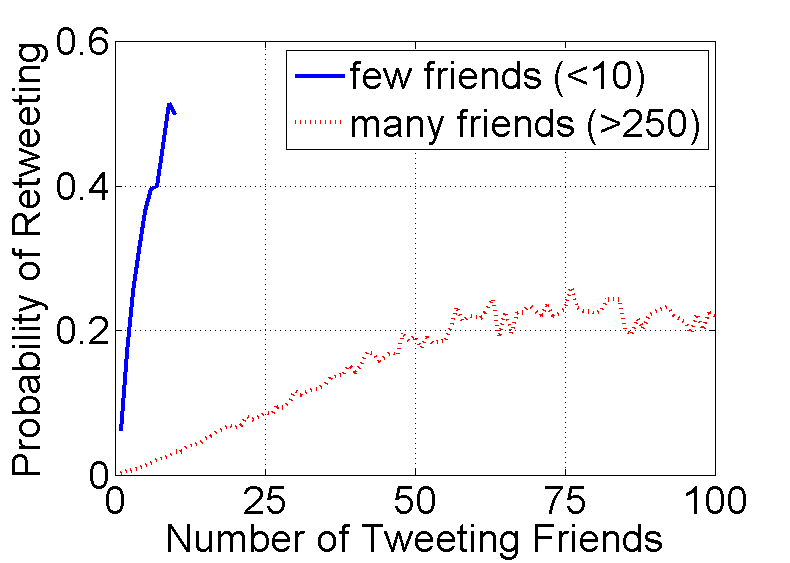}
\\
(a) & (b)
\end{tabular}
\caption{\emph{Exposure response in social media.} The probability to retweet some information as a function of the number of friends who previously tweeted it has a nonmonotonic trend when averaged over all users (a), but increases monotonically when users are separated according to the number of friends they follow (b). This suggests that additional exposures increase retweet likelihood, instead of suppressing it.
}
 \label{fig:twitter}
\end{figure*}

When examining how users spread information on a social media site Twitter, it may appear that repeated exposures to hashtags or links to online content make an individual less likely to use the hashtag himself or herself (Figure 1 of \cite{Romero11www}) or share the links with followers~\cite{Versteeg11icwsm} (Fig.~\ref{fig:twitter} (a)). From this, one may conclude the additional exposures ``inoculate'' the user and suppress the sharing of information. In fact, the opposite is true: additional exposures monotonically increase the user's likelihood to share information with followers~\cite{Lerman2016futureinternet}. The paradox arises because those users who follow many others---and are likely to be exposed to information or a hashtag multiple times---are less responsive overall (Fig.~\ref{fig:twitter} (b)), simply because they are overloaded with a large volume of information they receive~\cite{Hodas12socialcom}. Calculating response as a function of the number of exposures in the aggregate data falls prey to survivor bias: the more responsive users (with fewer friends) quickly drop out of analysis (since they are generally exposed fewer times), leaving only the highly connected, but less responsive users behind.  Their reduced susceptibility biases aggregate response, leading to wrong conclusions about individual behavior. Once data is disaggregated based on the volume of information individuals receive~\cite{Rodriguez14quantifying}, a clearer pattern of response emerges, one that is more predictive of behavior~\cite{Hodas14srep}.

\paragraph{Content Consumption in Social Media.}
\begin{figure*}[t]\centering
\begin{tabular}{cc}
\includegraphics[height=1.3in]{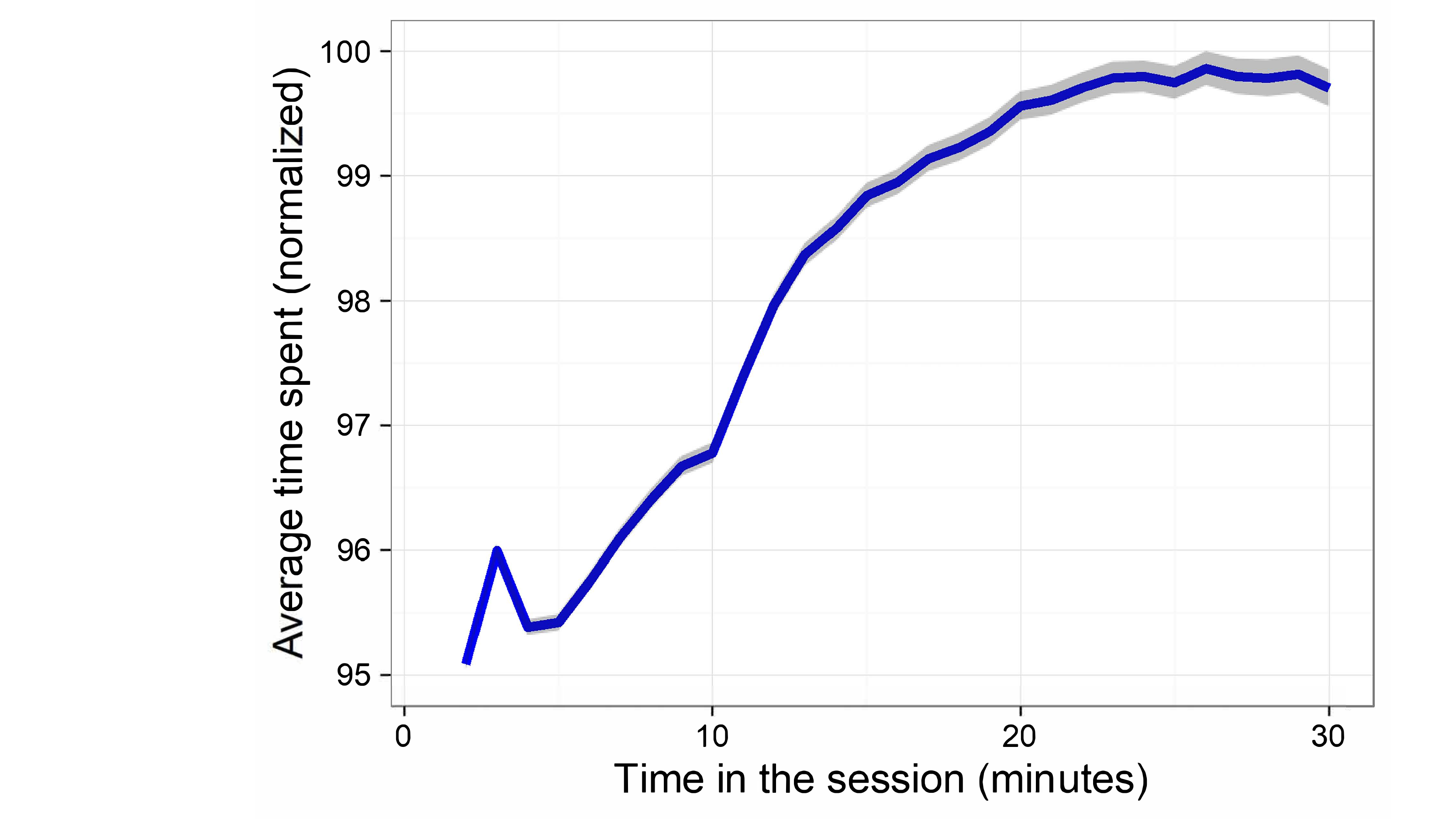}
\includegraphics[height=1.3in]{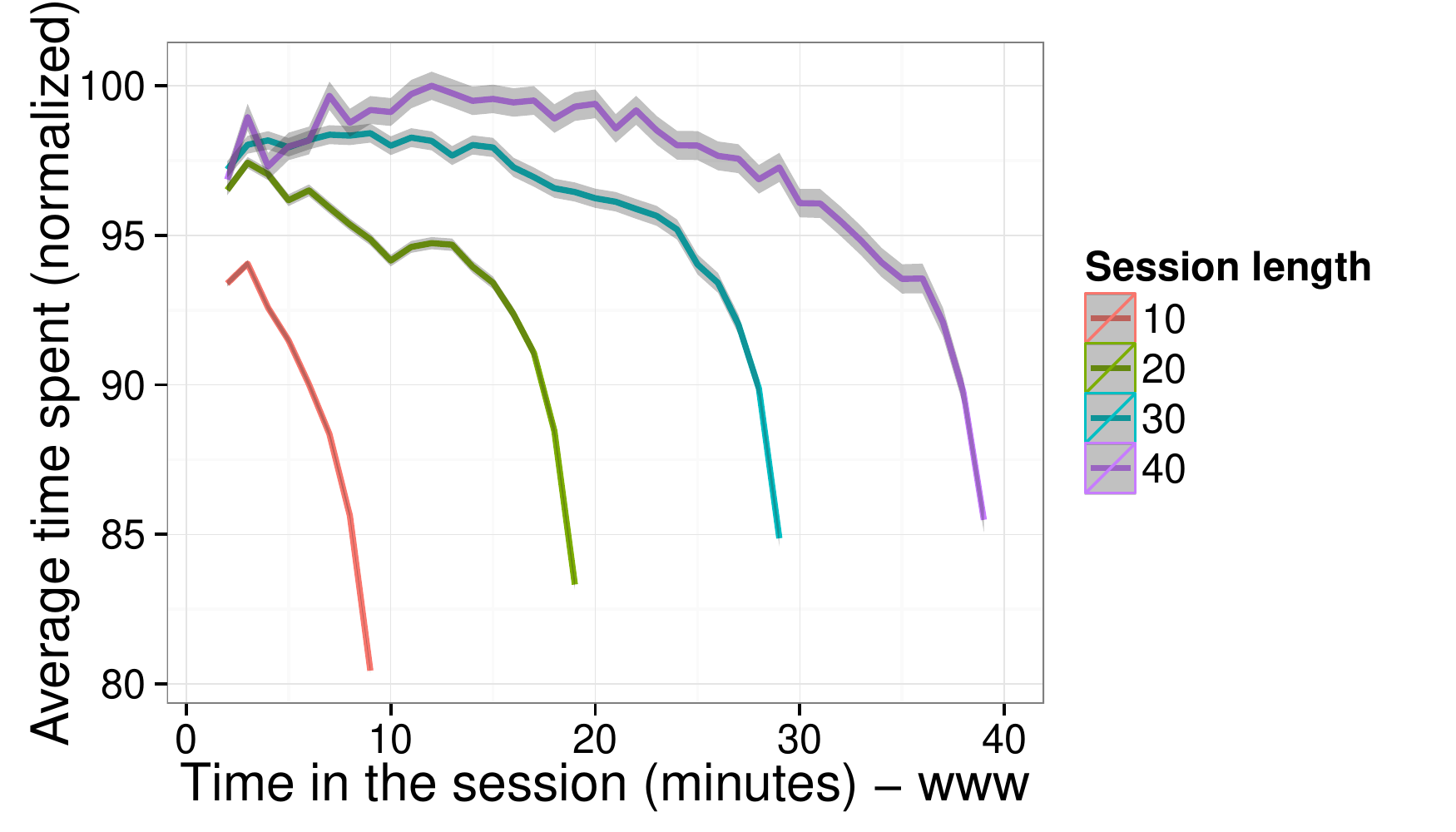}
\\
(a) & (b)
\end{tabular}
\caption{\textit{Rate of content consumption during a session.}
Average time spent viewing each item in a social feed appears to increase over the course of a session when looking at all the data (a) but decreases within sessions of the same length (b). This indicates that users speed up near the end of the session, taking less and less time to view each item.
}
\label{fig:fb}
\end{figure*}

A study of content consumption on a popular social networking site Facebook examined the time users devote to viewing each item in their social feed~\cite{KootiA2017www}. The study segmented each user's activity into sessions, defined as sequences of activity without a prolonged break (see Fig.~\ref{fig:randomization} for an explanation). At a population level, it looks as if users slow down over the course of a session, taking more and more time to view each item (Fig.~\ref{fig:fb} (a)). However, when looking at user activity within sessions of the same length, e.g., sessions that are 30 minutes long, it appears that individuals speed up instead (Fig.~\ref{fig:fb} (b)). As the session progresses, they spend less and less time viewing each item, which suggests that they begin to skim posts.

The difference in trends arises because users who have longer sessions also tend to spend more time viewing each item in their feed. When calculating how long users view items as a function of time, the faster users drop out of analysis of aggregate data, leaving the slower, users who tend to have longer sessions. Therefore, stratifying data by session length removes the confounding factor and allows us to study behavior within a similar cohort.

%

\paragraph{Answer Quality on Stack Exchange.}

\begin{figure*}[t]\centering
\begin{tabular}{cc}
\includegraphics[height=1.5in]{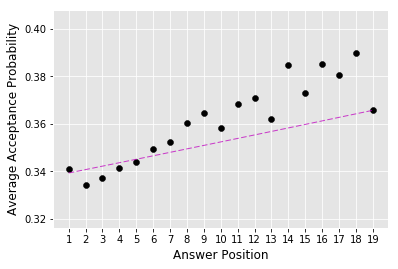}
\includegraphics[height=1.5in]{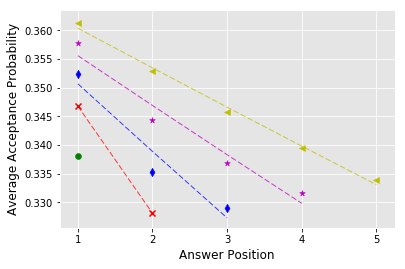}
\\
(a) & (b)
\end{tabular}
\caption{\textit{Quality of answers on Stack Exchange.} Probability that an answer is accepted as the best answer to a question increases as a function of its position within the session in the aggregated data (a) but decreases within sessions of the same length (b). This suggests that the quality of answers written by users deteriorates over the course of a session. Note that each line in the right panel represents sessions of a given length. Only sessions with five or fewer answers are shown.}
\label{fig:se}
\end{figure*}

Stack Exchange is a popular question-answering platform where users ask and answer questions. Askers can also ``accept'' an answer as the best answer to their question. A study of dynamics of user performance on Stack Exchange found that answer quality, as measured by the probability that it will be accepted by the asker as the best answer, declines steadily over the course of a session, with each successive answer written by a user ever less likely to get accepted~\cite{Ferrara2017dynamics}. However, this trend is seen only when comparing sessions of the same length, for example, sessions where exactly four answers were written (Fig.~\ref{fig:se} (b)). When calculating answer acceptance probability over all the data, it looks as though answers written later in a session are more likely to get accepted (Fig.~\ref{fig:se} (a)). Here, the length of the session confounds analysis: users who have longer sessions write answers that are more likely to be accepted.

%
%
%
%
%

\section{Testing Data for Simpson's Paradox}

\begin{figure}[t!]
	\centering
	\includegraphics[height=2in]{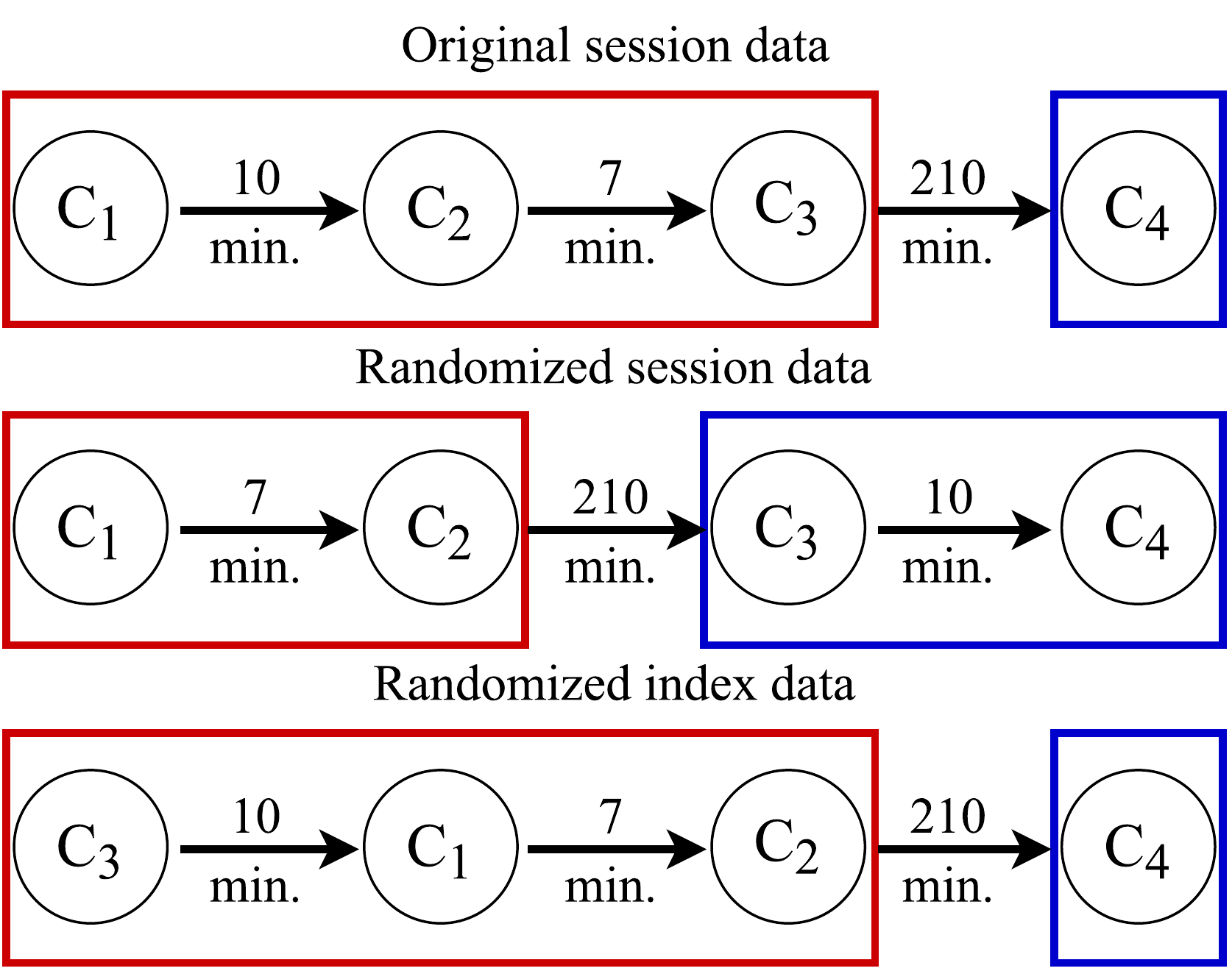}
    \label{shuffling}
	\caption{\emph{Data randomization for the shuffle test.} The top row shows the original stream of user actions $C_1, \ldots, C_4$. A session is a sequence of actions without an extended break, e.g., 60 minutes. Here, user actions $C_1$ through $C_3$ are assigned to one session, while $C_4$ is assigned to a new session. The middle row shows  data randomization strategy that shuffles time intervals between actions while preserving their order. This tends to change the definition of sessions. The bottom row shows the second randomization strategy, which shuffles the order of actions within sessions, while preserving the time intervals between actions.}
	\label{fig:randomization}
\end{figure}

When can a cautious researcher accept results of analysis? I describe a simple test that can help ascertain whether a pattern observed in data is robust or potentially a manifestation of Simpson's paradox. The test creates a randomized version of the data by shuffling it with respect to the  attribute for which the trend is measured. Shuffling preserves the distribution of features, but destroys correlation between the outcome variable and that attribute.
As a result, any trends with respect to that attribute should disappear.  This suggests a rule of thumb: if the trend persists in the aggregate data, but disappears when the shuffled data is disaggregated, then Simpson's paradox may be present.

In the analyses described above, the independent variable was time, or a proxy of it, such as the point within a session when the action takes place.
There are at least two different randomization strategies with respect to time. The first strategy creates \emph{randomized session data} by preserving the temporal order of actions, but shuffling the time intervals between them, as shown in Fig.~\ref{fig:randomization} (middle row). Since session break is defined as a sufficiently long time interval between actions, shuffling time intervals will merge sessions and break up longer sessions, while preserving the sequence of actions.
The second strategy creates a \emph{randomized index data} by shuffling the order of actions within a session, e.g., exchanging $C_1$ by $C_3$ in Fig.~\ref{fig:randomization} (bottom row).

Below I illustrate the shuffle test with real-world examples. I show that when the data is shuffled, the trend still persists in the aggregate data, but disappears, as expected, when the shuffled data is disaggregated.

\paragraph{Online Shopping.}
\begin{figure*}
\centering
\begin{tabular}{cc}
\includegraphics[height=1.5in]{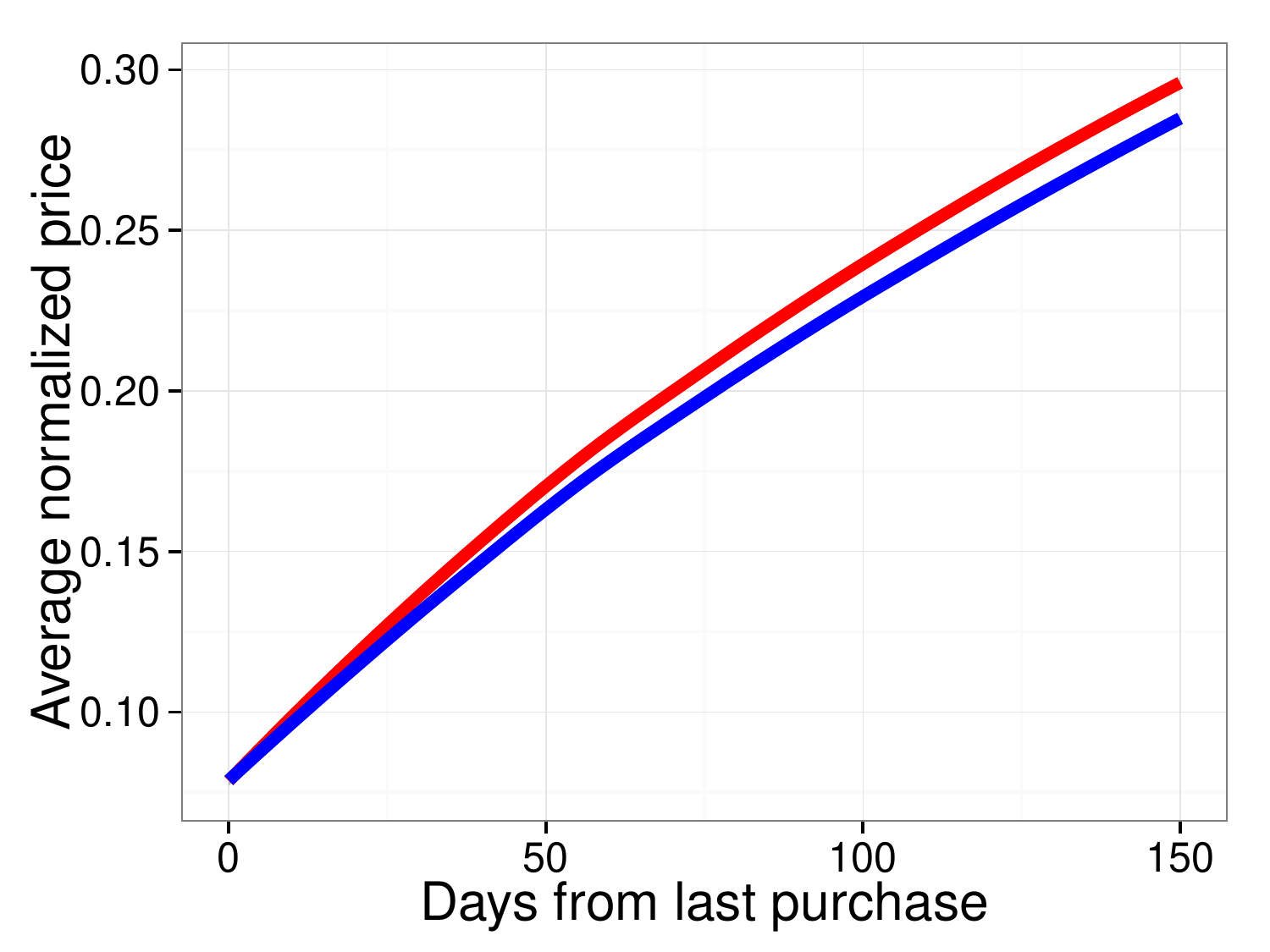}
\includegraphics[height=1.5in]{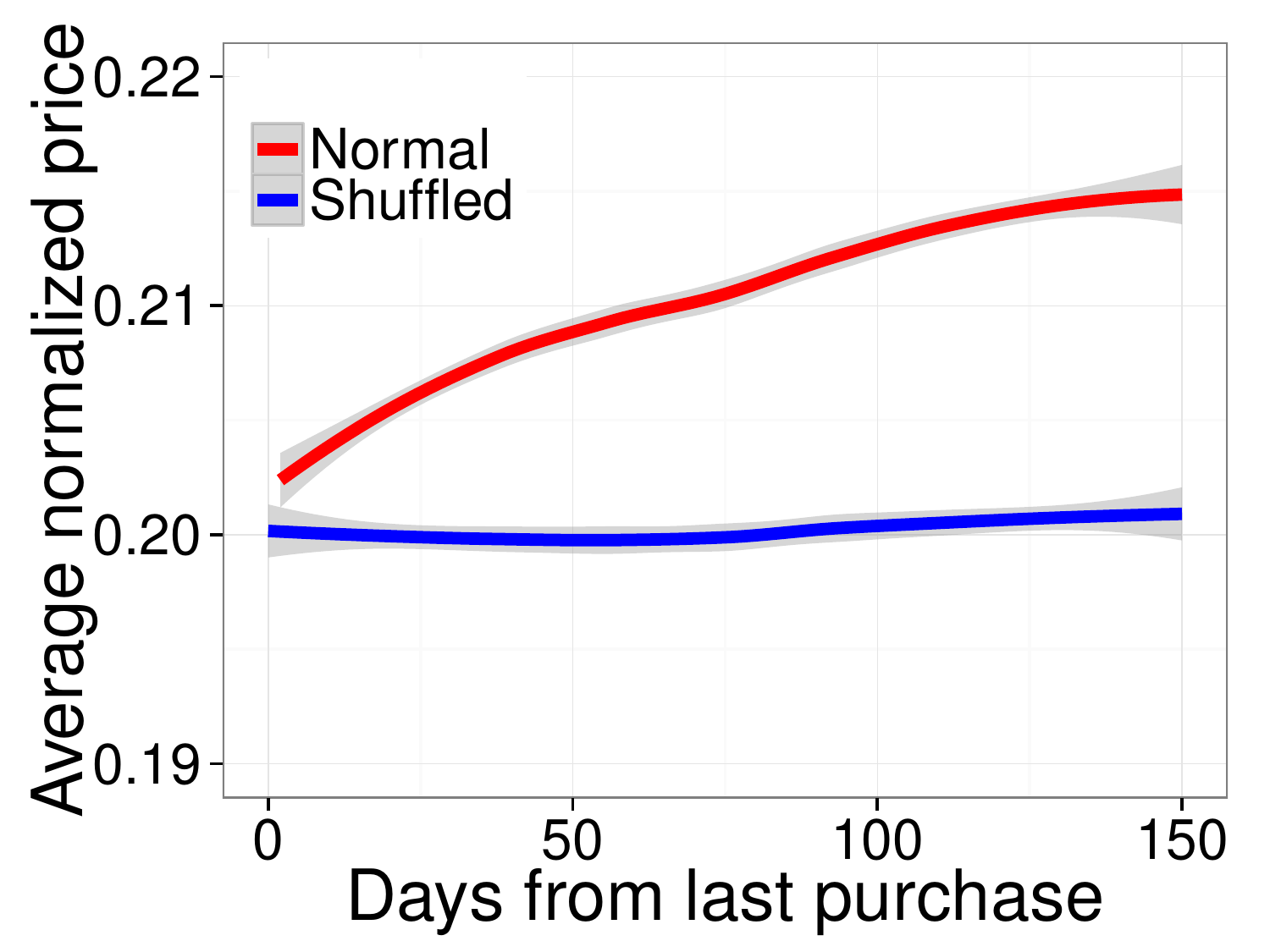}
\\
(a) & (b)
\end{tabular}
\caption{\textit{Online shopping.} Relationship between purchase price and time to next purchase in data (red line) and in the shuffled data (blue line), in which the purchase prices of items were randomly shuffled.
The positive trend seen in the aggregate data (a) still persists when data is shuffled. However, when data is disaggregated by the number of purchases, specifically, users who made exactly five purchases (b), the trend disappears in the shuffled data. }
\label{fig:online-shopping}
\end{figure*}

A study of online shopping examined whether individual purchasing decisions are constrained by finances. The study looked at the relationship between purchase price of an item and the time interval since last purchase~\cite{Kooti16wsdm}. Budgetary constraints would force a user to wait after making a purchase to accumulate enough money for another purchase. Figure~\ref{fig:online-shopping} (a) reports (normalized) purchase price of an item as a function of the time since last purchase (red line). The longer the delay, the larger the fraction of the budget users spend on a purchase, which appears to support the hypothesis.

To test the robustness of this finding, the data was shuffled by randomly swapping the prices of products purchased by users, which destroys the correlation between the time between purchases and purchase price. Surprisingly, the trend remains (blue line).
This is due to heterogeneity of the underlying population: the population represents a mix of users with different purchasing habits. The frequent buyers purchase  cheaper items more frequently, and they are systematically overrepresented on the left side of the plot, even in shuffled data.

To stratify data, buyers were grouped by the number of purchases they make, for example, those making exactly five purchases (Fig.~\ref{fig:online-shopping} (b)). 
The positive trend between the normalized purchase price and time seen in the disaggregated data (red line) disappears in the shuffled data (blue line), giving unbiased support for the limited budget hypothesis.

\paragraph{Stack Exchange.}
\begin{figure*}
\centering
\begin{tabular}{cc}
\includegraphics[height=1.5in]{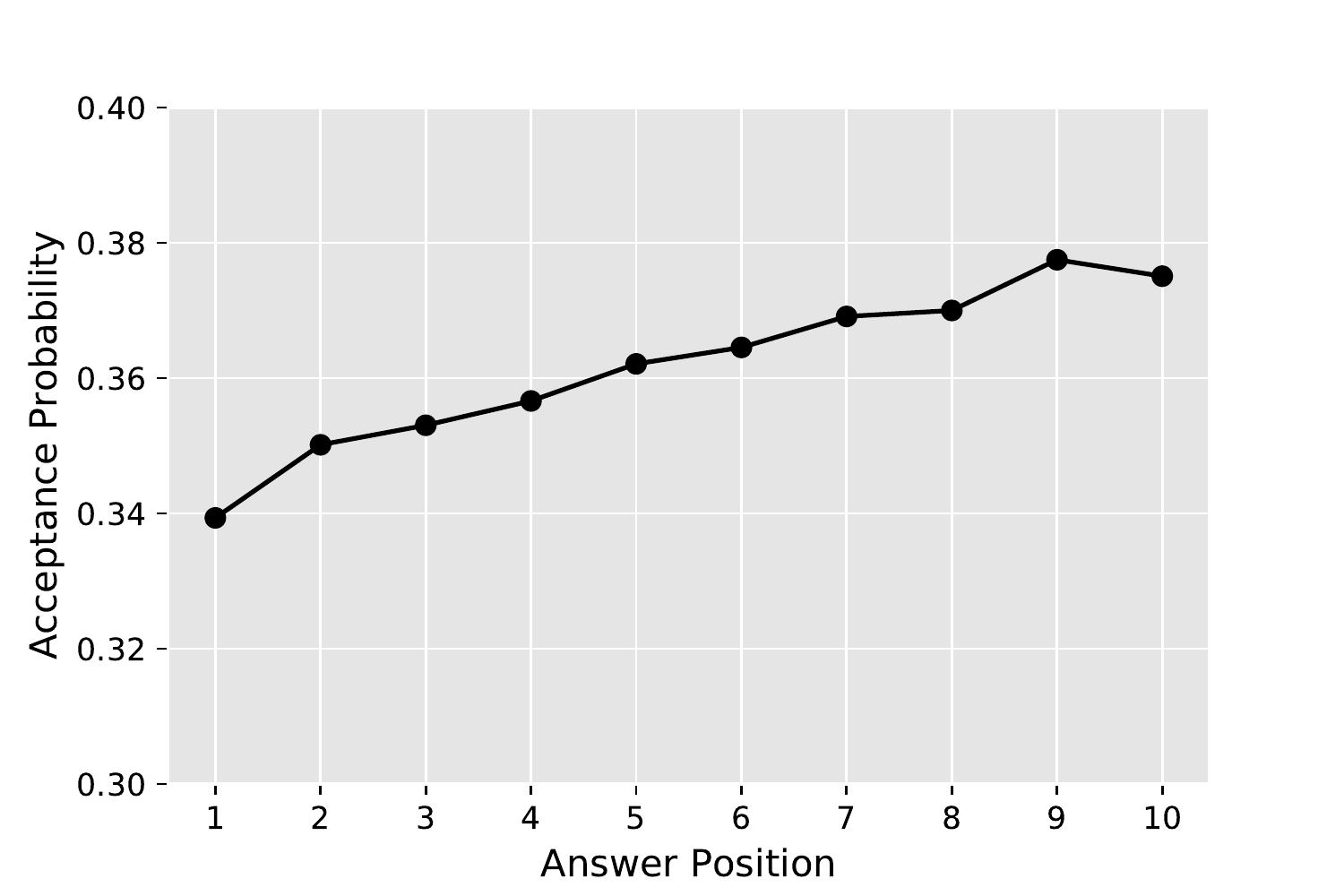}
\includegraphics[height=1.5in]{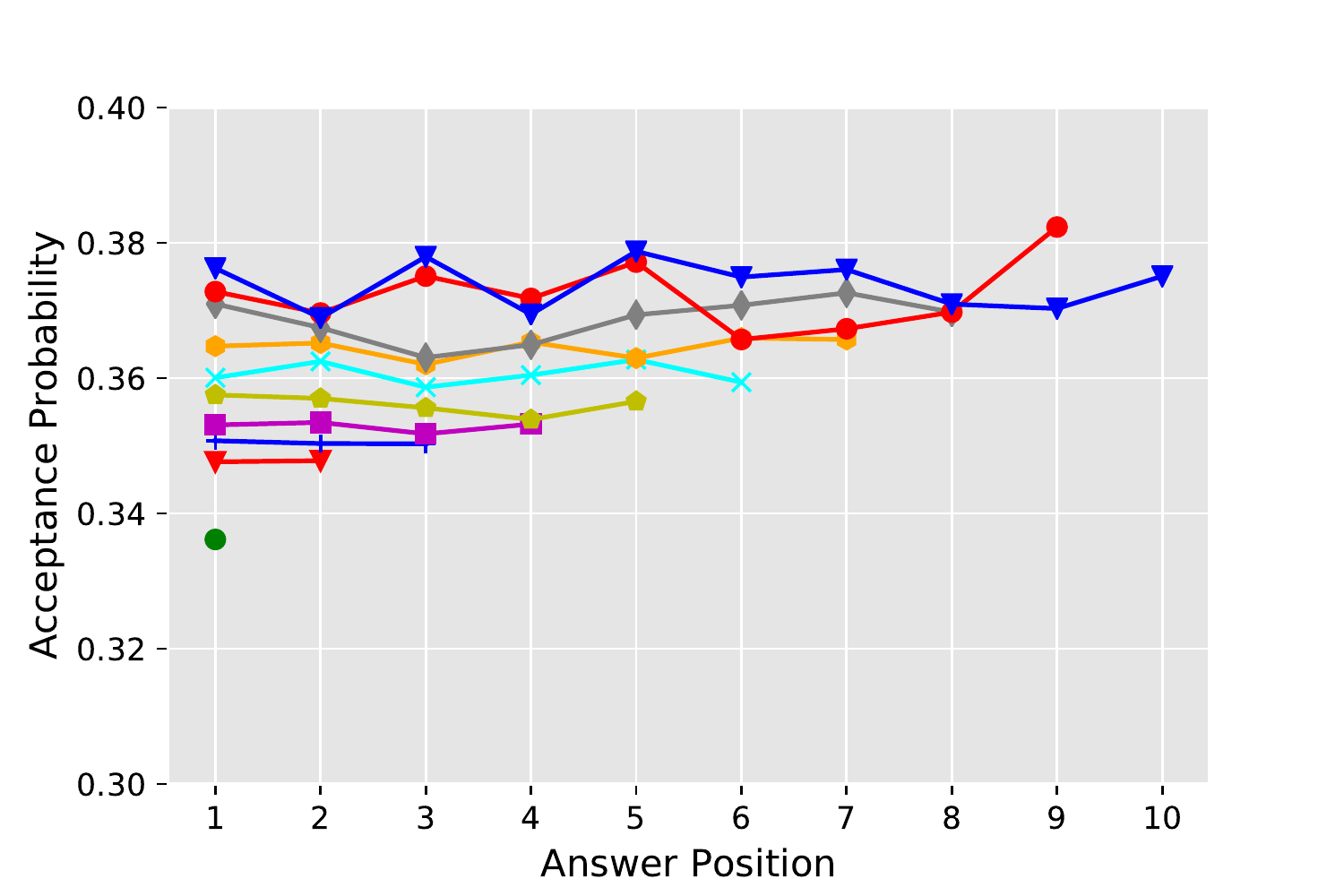}
\\
(a) & (b)
\end{tabular}
\caption{\textit{Answer's acceptance probability as a function of its session index in the randomized Stack Exchange data.} The left panel shows that the upward trend seen in Fig.~\protect\ref{fig:se} is preserved in the aggregate shuffled data. However, when shuffled data is disaggregated by session length (b), the trends largely disappear. }
\label{fig:se-randomized}
\end{figure*}

To test robustness of trends shown in Figure~\ref{fig:se}, which reports how acceptance probability of an answer posted on Stack Exchange changes over the course of a session, we randomize data by shuffling the time intervals between answers posted by each user, while preserving other features, including the temporal order of answers. The randomization procedure changes sessions by breaking up longer sessions and concatenating shorter ones. By changing which sequence of answers is  considered to belong to a session, we expect randomization to change the observed trends in acceptance probability.

The upward trend in acceptance probability seen in aggregate data still exists in the randomized data (Fig.~\ref{fig:se-randomized} (a)), even though the trends in randomized data  disappear, as expected, when data is disaggregated by session length (Fig.~\ref{fig:se-randomized} (b)).
This confirms the need for stratifying data by session length in analysis.

\paragraph{Reddit Comments.}
\begin{figure*}[h!t!]
	\centering
	\textbf{Original session data}\\
{\includegraphics[width=\textwidth]{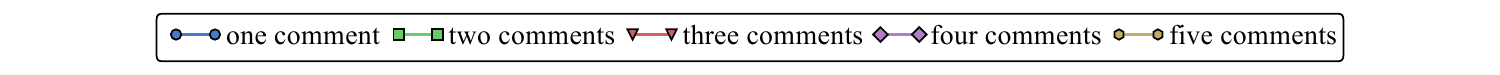}}
    \includegraphics[height=1.2in]{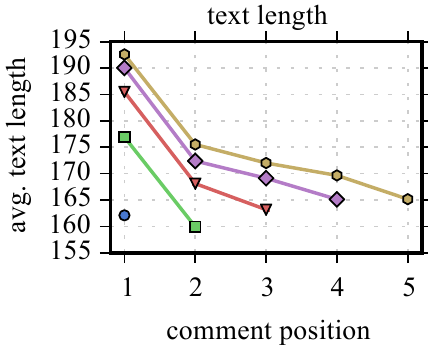}
	\includegraphics[height=1.2in]{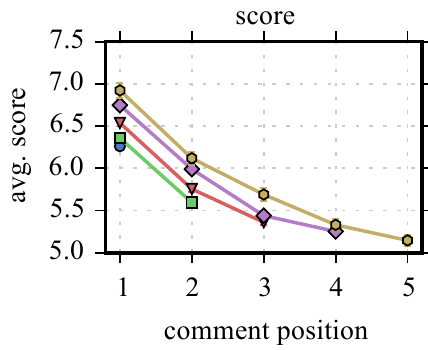}
	\includegraphics[height=1.2in]{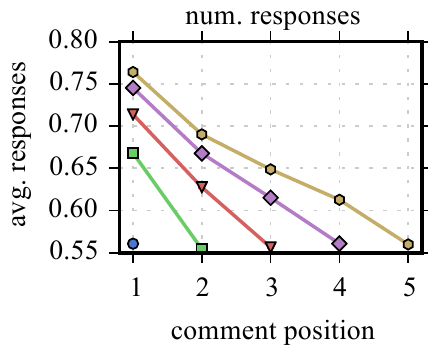}
\\
	\textbf{Randomized session data} \\
{
    \includegraphics[height=1.2in]{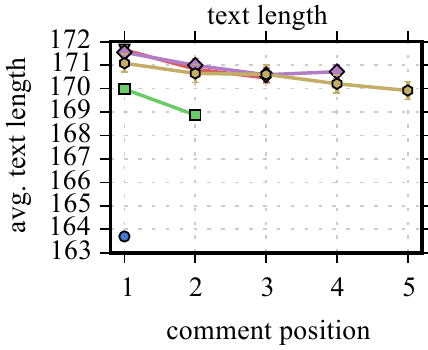}
	\includegraphics[height=1.2in]{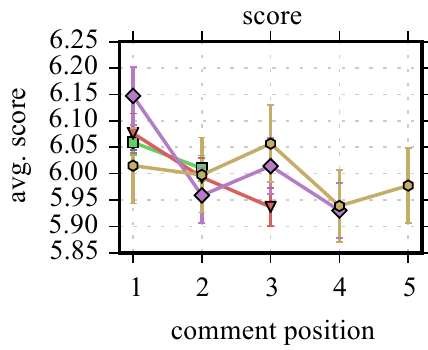}
	\includegraphics[height=1.2in]{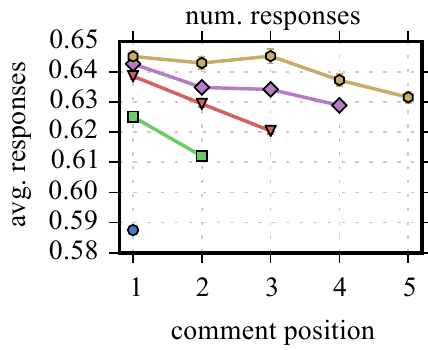}
	}
	\caption{
\emph{Deterioration in comment quality on Reddit.} When data is disaggregated by length of the session (different color lines), the quantitative proxies of comment quality decline over the course of a session. The x-axis represents index of the comment within a session, and the y-axis gives the average value of the proxy measure (with error bars). The declines observed in original Reddit data (top row) mostly disappear when data is randomized (bottom row).
}
	\label{fig:reddit-randomized}
\end{figure*}

A similar quality deterioration effect was observed for comments posted on Reddit. Regardless of what measure is used as a proxy of quality---comment length, the number of responses or upvotes from others it receives, its textual complexity---the quality of each successive comment written by a Reddit user decreases over the course of a session~\cite{Singer2016plosone}. To test the robustness of this finding, Singer et al. randomized Reddit activity data.
Figure~\ref{fig:reddit-randomized} compares the trends for the proxies of comment quality in the original data to those in the randomized data. Both data sets have been disaggregated by session length. 
The decreasing trends observed in the original Reddit data (top row) largely disappear in the randomized data (bottom row). Where the trends still exist, the deterioration effect is much reduced. This suggests that most of data heterogeneity is captured by session length.

\section{Conclusion}
Simpson's paradox can indicate that interesting patterns exist in data~\cite{fabris2000discovering}, but it can also skew analysis. The paradox suggests that data comes from subgroups that differ systematically in their behavior,  and that these differences are large enough to affect analysis of aggregate data. 
In this case, the trends discovered in disaggregated data are more likely to describe---and predict---individual behavior than the trends found in aggregate data. Thus, to build more robust models of behavior, computational social scientists need to identify confounding variables which could affect observed trends. The shuffle test described in this paper provides a framework for determining whether Simpson's paradox is affecting conclusions.


\subsection*{Acknowledgements}
Many people have contributed along the way to identifying the problem of Simpson's paradox in data analysis, investigating it empirically, as well as devising methods to mitigate its effects. These people include Nathan Hodas, Farshad Kooti, Keith Burghardt, Philipp Singer, Emilio Ferrara, Peter Fennell, Nazanin Alipourfard.
This work was funded, in part, by Army Research Office under contract W911NF-15-1-0142.



\end{document}